# Interplay of Rotational, Relaxational, and Shear Dynamics in Solid $^4$He

E.J. Pratt,[1,2§] B. Hunt,[1,3§] V. Gadagkar,[1] M. Yamashita,[4] M. J. Graf[5], A. V. Balatsky[5] and J.C. Davis[1,6,7]

[1]Laboratory for Atomic and Solid State Physics, Department of Physics, Cornell University, Ithaca, NY 14853, USA. [2]Kavli Institute for Theoretical Physics, UC Santa Barbara, CA 93016, USA. [3]Department of Physics, Massachusetts Institute of Technology, Cambridge MA, 02139. [4]Department of Physics, Kyoto University, Kyoto 606-8502, Japan. [5]Theoretical Division and Center for Integrated Nanotechnology, Los Alamos National Lab., Los Alamos, NM 87545, USA. [6]Condensed Matter Physics and Materials Science Department, Brookhaven National Laboratory, Upton, NY 11973, USA. [7]School of Physics and Astronomy, University of St. Andrews, St. Andrews, Fife KY16 9SS, UK.
[§]These authors contributed equally to this work.

Using a high-sensitivity torsional oscillator (TO) technique, we map the rotational and relaxational dynamics of solid $^4$He throughout the parameter range of the proposed supersolidity. We find evidence that the same microscopic excitations controlling the torsional oscillator motions are generated independently by thermal and mechanical stimulation. Moreover a measure for the relaxation times of these excitations diverges smoothly without any indication for a critical temperature or critical velocity of a supersolid transition. Finally, we demonstrate that the combined temperature-velocity dependence of the TO response is indistinguishable from the combined temperature-strain dependence of the solid's shear modulus. This implies that the rotational responses of solid $^4$He attributed to supersolidity are associated with generation of the same microscopic excitations as those produced by direct shear strain.

Solid $^4$He may become a supersolid (*1*) when its temperature $T$ and mass-flow velocity $V$ fall below their critical (*2*) values $T_c$ and $V_c$. Indeed, torsional oscillator (TO) studies ( *3*, *4* ) reveal that the resonant angular frequency of rotation $\omega$ increases rapidly below both $T$~250 mK and rim-velocity $V$~$10^{-4}$ ms$^{-1}$, as if superfluid inertia decouples at a critical temperature and velocity. And these $\omega_0$ increases (*3- 10*) are greatly diminished by blocking the TO annulus (*4*, *11*), as if superfluid inertia is thereby reconnected. Signatures in the heat capacity ascribed to supersolidity also occur in this same temperature range (*12*). However, direct mass-flow studies detect maximum currents which are far smaller than those implied by the TO experiments (*13-15*). Moreover, the temperature dependence of the resonance frequency $f(T)=\omega(T)/2\pi$ of TOs containing solid $^4$He (*3 - 11*) resembles closely that of its shear modulus $\mu(T)$ (*16*). And, coincident with the maximum rates of increase of $f(T)$ and $\mu(T)$ are maxima in TO dissipation (*4-6*, *8*, *9*) and shear-dissipation (*16, 17*), respectively. Such effects should not exist during a bulk Bose-Einstein condensation transition (although they do occur in the Berezinskii-Kosterlitz-Thouless transition of a superfluid film (*18*) - see Supplementary Online Material (SOM) section I (19) ). Finally, the increases in both $f$ and $\mu$ are quickly extinguished by increasing TO maximum rim-velocity $V$ (*3-8*, *10*) or shear strain (*16, 20*) $\varepsilon$, respectively.

Several theoretical models have been proposed to explain the unexpectedly complex rotational dynamics of solid $^4$He. The first is a simple supersolid (*1*) in which Bose-Einstein condensation of vacancies produces an interpenetrating superfluid with well defined $T_c$ and $V_c$. The second is an incipient supersolid lacking long range phase coherence (*21*, *22*). A third class of model posits disorder-induced superfluidity (*9*, *23- 29*). The final proposal is that solid $^4$He contains a population of inertially active crystal excitations (*30- 35*) whose relaxation time $\tau$ lengthens smoothly with falling $T$ and $V$. These excitations are variously proposed to be a dynamical network of pinned dislocations (*17,30,35*), atomic-scale tunneling two-level systems (34), and the glassy response of defects distributed throughout the solid (*31-33*). All models positing inertially active crystal excitations have the property that, as $\tau$(T) passes through the condition $\omega\tau$=1, a strong maximum in $|df/dT|$ and TO dissipation $D$ should occur (*9, 31-35*) even though there is no supersolid $T_c$ and $V_c$. By contrast, a bulk superfluid phase transition should exhibit clear signatures of both $T_c$ and $V_c$ (2). One way to distinguish between such models is to determine the evolution of microscopic relaxational time-constants $\tau$, in search of either the smoothly diverging $\tau$ of a system governed by $\omega\tau$=1 phenomenology or the sudden changes expected in $\tau$ at a thermodynamic $T_c$ and/or $V_c$.

An unbiased approach to TO studies of solid $^4$He can be achieved by using the TO rotational susceptibility $\chi(\omega,T)=\theta(\omega,T)/\Gamma(\omega)$ (*9*). Here $\theta(\omega,T)$ represents the amplitude of angular dispalcement as a function of $\omega$ and $T$ in response to a harmonic torque $\Gamma(\omega)$ of constant magnitude. Then(*31, 32, 33*)

$$\chi^{-1}(\omega,T)=K-I\omega^2-i\gamma\omega-\chi^{-1}_{4He}(\omega,T) \qquad (1)$$

represents the the properties of the bare TO plus the 'back action' of the solid $^4$He upon it through the solid's rotational susceptibility $\chi^{-1}_{4He}(\omega,T)$. Here $I$ is the combined moment of inertia of the TO plus $^4$He at zero-temperature, $K$ is the torsional spring constant, and $\gamma$ is the TO damping constant. To clarify these concepts, we consider a Debye rotational susceptibility $\chi^{-1}_{4He}(T)=g/(1-i\omega_0\tau(T))$ (*9, 31 -33*) with relaxational time constants $\tau(T)$ increasing with decreasing $T$. For this susceptibility

$$\frac{\mathrm{Im}(\chi^{-1}_{4He})}{\mathrm{Re}(\chi^{-1}_{4He})}=\frac{D(T)f(0)}{2(f(0)-f(T))}=\omega_0\tau(T) \qquad (2)$$

where $\mathrm{Re}(\chi^{-1}_{4He})$ and $\mathrm{Im}(\chi^{-1}_{4He})$ are its real and imaginary parts respectivly (*32*), and $D(T)=Q^{-1}(T)-Q^{-1}(T\to\infty)$ is the inverse contribution to the TO quality factor $Q$ from the solid $^4$He. Access to $\tau(T)$ for the microscopic excitations is therefore possible in principle from measurements of $\mathrm{Re}(\chi^{-1}_{4He})$ and $\mathrm{Im}(\chi^{-1}_{4He})$ .

Following this approach, we map the rotational susceptibility of a TO containing solid [4]He throughout the $V$-$T$ plane (SOM section II). The results in Figs 1A,B reveal immediately that the frequency increase and dissipation peak are bounded by closely corresponding $V$-$T$ contours. Thus the same unexplained dissipation seen with falling temperature near the proposed supersolid $T_c$ is found also with diminishing $V$ in the range of the proposed (*3*, *4*) supersolid $V_c$. The highly similar contours of both $f(T,V)$ and $D(T,V)$ also reveal that the maxima in $|df/dT|$ and $D$ are always linked - as if controlled by some combined function of $T$ and $V$. Similar results were observed in all three distinct solid [4]He samples studied.

Next we compare the solid [4]He rotational dynamics versus $T$ as $V \rightarrow 0$ to those versus $V^{0.5}$ as $T \rightarrow 0$ (the rationale for $V^{0.5}$ will become clear below). Figures 2A and 2C show $f(T)|_{V\rightarrow 0}$ and $D(T)|_{V\rightarrow 0}$ while Figs 2B and 2D show $f(V)|_{T\rightarrow 0}$ and $D(V)|_{T\rightarrow 0}$ (Fig. 1 data used are identified in SOM section III, Fig. S5). Figure 2 reveals a striking and unexpected similarity between the results of what, for a simple superfluid, would be two completely different experiments (one stimulating the sample thermally and the other mechanically). To examine this, we define an empirical measure $\tau_E$ of relaxation times for any combination of $T$ and $V$:

$$\tau_E(T,V) = \frac{D(T,V)\,f(0)}{2\omega_0(f(0) - f(T,V))} \qquad (3)$$

In Fig. 2E we show $\log \tau_E(T)$ plotted versus $\log(T/T^*)$ for the lowest rim-velocity data (Figs 2A,C). In Fig. 2F, $\log \tau_E(V)$ is likewise plotted versus $\log(V/V^*)$ for the lowest temperature data (Figs 2B,D). Here we define $T^*$ and $V^*$ as the temperature and rim-velocity respectively at which half the total frequency shift has occured (Figs 1, 2A, 2B). This analysis reveals that the $\tau_E$ diverge smoothly as $T^{\zeta}$ with $\zeta = -2.75 \pm 0.1$ when $V \rightarrow 0$, and as $V^{\lambda}$ with $\lambda = -1.17 \pm 0.05$ when $T \rightarrow 0$. Thus, the effects of temperature on $f(T)|_{V\rightarrow 0}$ and $D(T)|_{V\rightarrow 0}$ appear identical to those of rim-velocity on $f(V^{\gamma})|_{T\rightarrow 0}$ and $D(V^{\gamma})|_{T\rightarrow 0}$ respectively, where $\gamma = \lambda / \zeta = 0.43$ is the ratio of power-law exponents. Figures 2E,F also shows that no matter how complex the actual rotational dynamics (Figs 2A-D), the peak in $D$ is always canceled by the peak in $|df/dT|$ to produce smoothly diverging functions $\tau_E(T)|_{V\to 0}$ and $\tau_E(V)|_{T\to 0}$ (see SOM section III, Fig. S7). Microscopic relaxational processes represented by $\tau_E$ should change dramatically at a superfluid phase transition; an excellent example of this is seen in $\tau_E(T)$ at the BKT superfluid phase transition of liquid $^4$He shown inset to Fig. 3E (see SOM section I, fig. S1). However, no indications of the sudden change which would signify the supersolid $T_c$ or $V_c$ exist in Figs 2E,F. Instead, $\tau_E$ exhibits everywhere the smooth divergence expected in $\omega\tau=1$ models.

Figures 1 and 2 provide direct empirical evidence that the effects of $T$ and $V$ on the TO are intimately related to each other. One may therefore ask whether a single Debye-like rotational susceptibility could describe the whole $V$-$T$ plane dynamics in Figure 1 when the effects of $V$ on the relaxation time $\tau$ are correctly incorporated. Hypothesizing a total relaxation rate $1/\tau(T,V)$ due to a combination of two effects

$$\frac{1}{\tau(T,V)}=\frac{1}{\tau(T)}+\frac{1}{\tau(V)} \quad (4)$$

along with the knowledge that the overall phenomenology appears identical as a function of $T^{\zeta}|_{V\to 0}$ and $V^{\lambda}|_{T\to 0}$ (Fig. 2) and interpolates smoothly between these limits (Fig. 1), yields an ansatz:

$$\frac{1}{\tau\ (T,V)}=\frac{\Sigma}{T^{\zeta}}+\frac{\Lambda}{V^{\lambda}} \quad (5)$$

Here $\Sigma$ and $\Lambda$ quantify the relative contributions to the relaxation rate from thermally and mechanically stimulated excitations (SOM section IV). Figure 3A shows that by using this ansatz, all the complex solid $^4$He rotational dynamics in $f(T,V)$ and $D(T,V)$ of Fig. 1 can be collapsed onto just two functions $\mathrm{Re}(\chi^{-1})$ and $\mathrm{Im}(\chi^{-1})$ merely by plotting $f[(T^*/T)^{\zeta}+(V^*/V)^{\lambda}]$ and $D[(T^*/T)^{\zeta}+(V^*/V)^{\lambda}]$. Moreover this apparent unification of rotational dynamics implies that Eqn. 3 could yield a comprehensive image of $\tau_E(T,V)$ throughout the $V$-$T$ plane by dividing all the data in Fig. 1B by that in Fig. 1A, as shown in Fig. 3B. Although the proposed *V-T* ranges for a supersolid phase transition (*3,4,12*) are at or below the dashed *T*-V** contour, the $\tau_E$ surface exhibits everywhere the smoothly diverging relaxation processes expected in $\omega_0\tau=1$ models. We emphasize here that all the above results (Figs 1-3) are independent of any choice of $\chi^{-1}_{4He}(T,V)$ and therefore strongly constrain eventual microscopic models for the dynamics of solid $^4$He.

Figures 2 and 3 provide evidence that the identical microscopic excitations are being generated by thermal and mechanical stimulation, and that the overall rotational dynamics in *f(V,T)* and *D(V,T)* are consistent with a single $\omega_0\tau=1$ mechanism that is controlled by a relaxation rate $(T^*/T)^{\zeta}+(V^*/V)^{\lambda}$ due to the combined influences from these two sources (Fig. 4). As these unified dynamics also appear inconsistent with expectations for $T_c$ or $V_c$ of a superfluid transition (2), one must ask which model could account for them? Because the solid $^4$He shear modulus $\mu(T)$ exhibits a very similar temperature dependence to *f(T)* (*16*), and because this shear stiffening effect is extinguished by a characteristic strain as opposed to a critical velocity (*20*), a key question has been whether excitations generated by direct shearing are the same as those controlling the TO dynamics.

Our approach provides a new opportunity to address this issue. If crystal excitations induced by inertial strain $\varepsilon$ in the TO (where $\varepsilon \propto V$) are the cause of the anomalous rotational dynamics, then the indistinguishable structure of $f(T)|_{V\to 0}$ and $f(V^{\gamma})|_{T\to 0}$ (Fig. 4A) should be mirrored by an equivalently indistinguishable relationship in shear modulus between $\mu(T)|_{\varepsilon\to 0}$ and $\mu(\varepsilon^{\gamma})|_{T\to 0}$. When the measured $\mu$ from Ref. 20 is plotted simultaneously versus *T* and $\varepsilon^{\gamma}$ in Fig. 4B (using the power-law ratio $\gamma$ derived from our TO studies), this is precisely what we find. That the combined temperature-velocity dependence of the TO response mirrors quantitatively the combined temperature-strain dependence of the shear modulus, along with the original observation that $\mu(T)$ tracks *f(T)* (*16*), implies that the rotational dynamics of solid $^4$He are associated with the generation (presumably by inertial shearing) of the same type of microscopic excitations which are generated by direct shear strain $\varepsilon$. These conclusions appear to be in good accord with the observed smoothly diverging microscopic relaxation times as expected of $\omega\tau=1$ models (Fig. 3), and with the absence of a signature in $\tau_E(T,V)$ for the $T_c$ or $V_c$ of a supersolid phase transition (Figs 2,3). These results will motivate efforts to: (i) identify directly whether the microscopic excitations are crystal dislocations as implied, and (ii) determine whether they admit any associated DC contribution to the rotational susceptibility which would represent a superfluid component (*9*).

**Supplementary Online Materials** are linked to the online version of the paper at www.sciencemag.org.

**Acknowledgements** We are grateful for discussions and communications with: J. Beamish, D. M. Ceperley, M. H. W. Chan, J. Day, A. T. Dorsey, R. B. Hallock, H. Kojima, D. M. Lee, A. J. Leggett, E. Mueller, D. R. Nelson, J. M. Parpia, N. V. Prokof'ev, J. D. Reppy, P. C. E. Stamp, B. Svistunov, and M. Troyer. These studies are supported by the National Science Foundation under grants DMR-0806629 and NSF PHY05-51164 (KITP); Work at Los Alamos was supported by U.S. Department of Energy grant DE-AC52-06NA25396 through LDRD.

**Figure 1.**

TO resonant frequency shift *f(T)* (A) and dissipation data (B) mapped throughout the V-T plane. Ninety eight FID curves (each at a different temperature) were smoothly interpolated into the two color-coded surfaces displayed here on identical log-log axes. The low-velocity maximum frequency shift (~33 mHz) would correspond to a superfluid fraction of 5.6% .

**Figure 2.**

A. TO resonant frequency shift *f(T)* measured at lowest rim-velocity. T* is defined as the temperature at which 50% of the frequency change has occurred (Fig 1).
B. TO resonant frequency shift *f(√V)* measured at lowest temperature . V* is defined as the rim-velocity at which 50% of the frequency change has occurred (Fig. 1).
C. TO dissipation *D(T)* measured at lowest rim-velocity .
D. TO dissipation *D(√V)* measured at lowest temperature .
E. The empirical measure of microscopic relaxation times $\omega_0 \tau_E(T)|_{V \to 0}$ from data in Figs 2A,C. The inset shows the equivalent analysis using Eq. 3 for the BKT transition of a superfluid $^4$He film (see SOM section I).
F. The empirical measure of microscopic relaxation times $\omega_0 \tau_E(V)|_{T \to 0}$ from data in Figs 2B,D . It diverges smoothly as $V^{\lambda}$ with $\lambda = -1.17 \pm 0.05$.

**Figure 3.**

**A.** All of the TO dynamical responses throughout the *V*-*T* plane (*f*(*T,V*) from Fig. 1A and *D*(*T,V*) from Fig. 1B) are collapsed onto just two curves (very similar in structure to the $\mathrm{Re}(\chi^{-1})$ and $\mathrm{Im}(\chi^{-1})$ components of the Debye susceptibility) by plotting $f[(T^*/T)^{\zeta}+(V^*/V)^{\lambda}]$ and $D[(T^*/T)^{\zeta}+(V^*/V)^{\lambda}]$ (SOM section IV). Here we find that $\mathrm{Re}(\chi^{-1})\propto f[(T^*/T)^{\zeta}+(V^*/V)^{\lambda}]$ is always too large in comparison to $\mathrm{Im}(\chi^{-1})\propto D[(T^*/T)^{\zeta}+(V^*/V)^{\lambda}]$ to be explained quantitatively by a Debye susceptibility model; this point has been used to motivate a "superglass" hypothesis $\mathrm{Re}(\chi^{-1})$ (*9*).

B. A comprehensive map of empirical relaxation times $\omega_0\tau_E(T,V)$ deduced using Eqn. 3 represented as a surface in the log*T*-log*V* plane. The equally-spaced contour lines in $\log\omega_0\tau_E(T,V)$ reveal the underlying divergence of $\omega_0\tau_E(T,V)$ as combined power laws $[(T^*/T)^{\zeta}+(V^*/V)^{\lambda}]$.

**Figure 4.**

A. Plots of our simultaneously measured $f(T)|_{V\to 0}$ (open circles) and $f(V^{\gamma})|_{T\to 0}$ (filled squares) from Figs 1 and 2.

B. Simultaneous plots of measured $\mu(T)|_{\varepsilon\to 0}$ (open circles) and $\mu(\varepsilon^{\gamma})|_{T\to 0}$ (filled squares) from Ref. 20 acquired at 2000 Hz.

Figure 1:

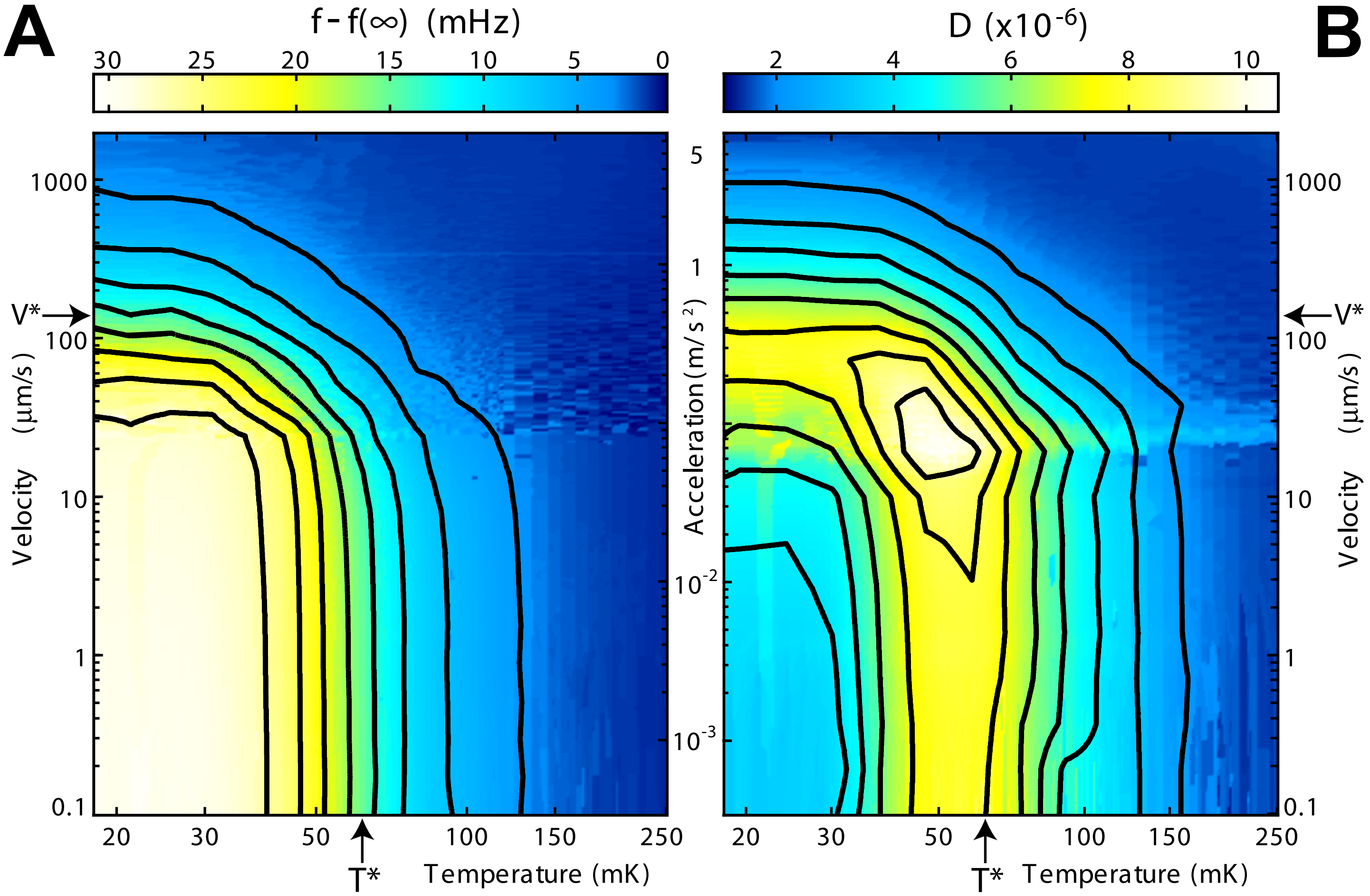

Figure 2:

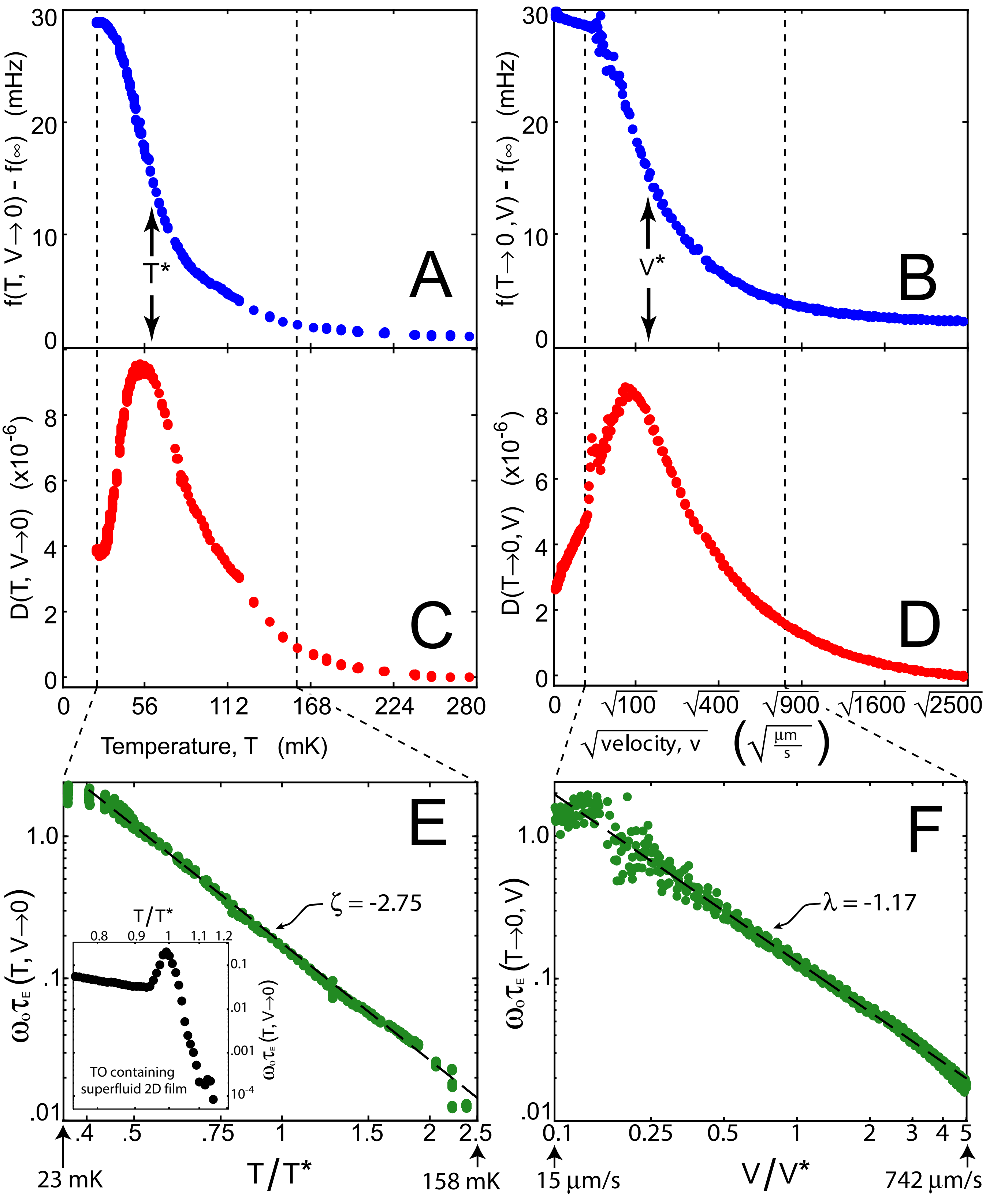

Figure 3:

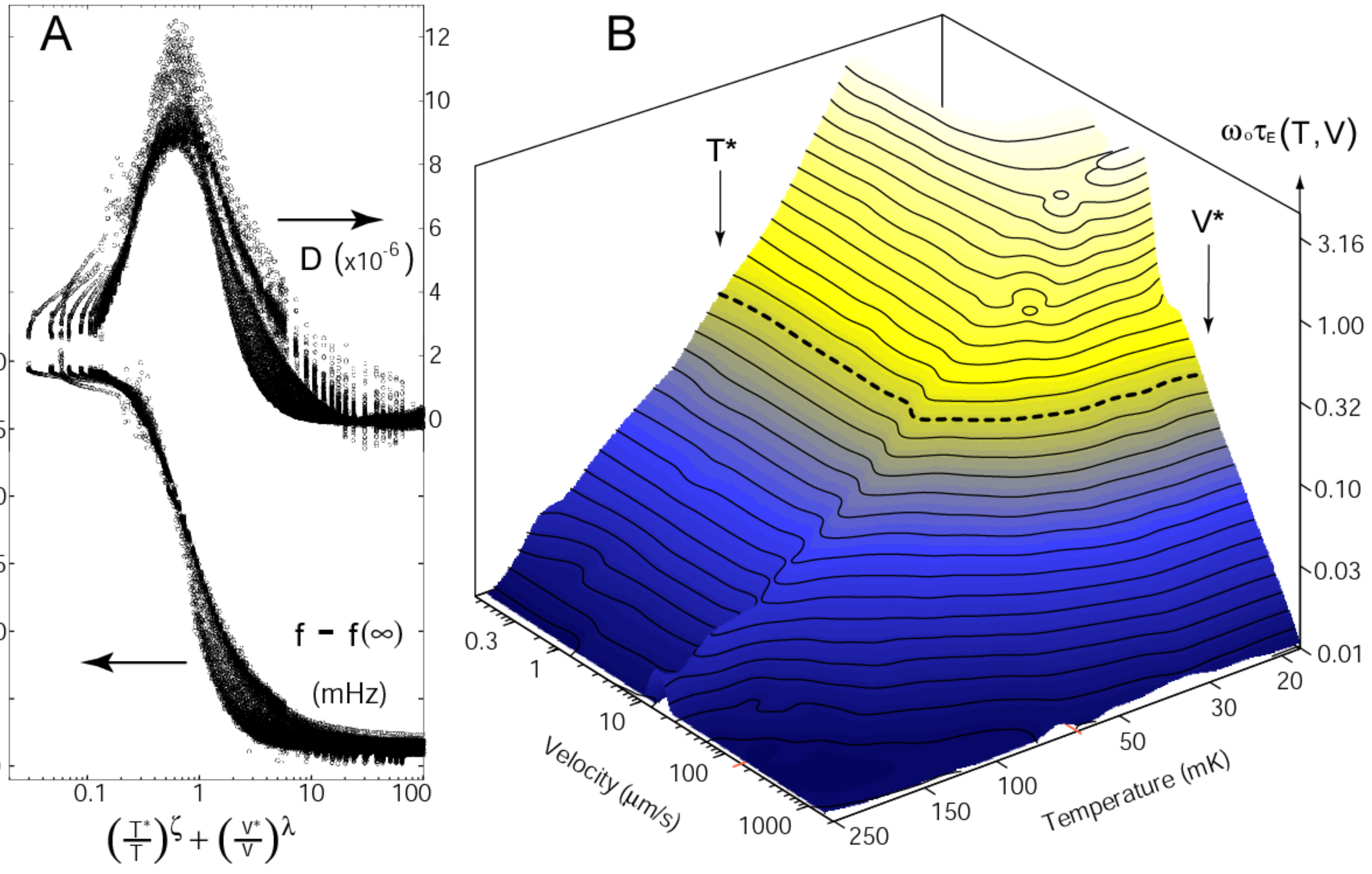

A
D (x10$^{-6}$)
f − f(∞)
(mHz)
$\left(\frac{T^*}{T}\right)^{\zeta} + \left(\frac{V^*}{V}\right)^{\lambda}$
B
T*
V*
$\omega_0\tau_E(T,V)$
Velocity (μm/s)
Temperature (mK)

Figure 4:

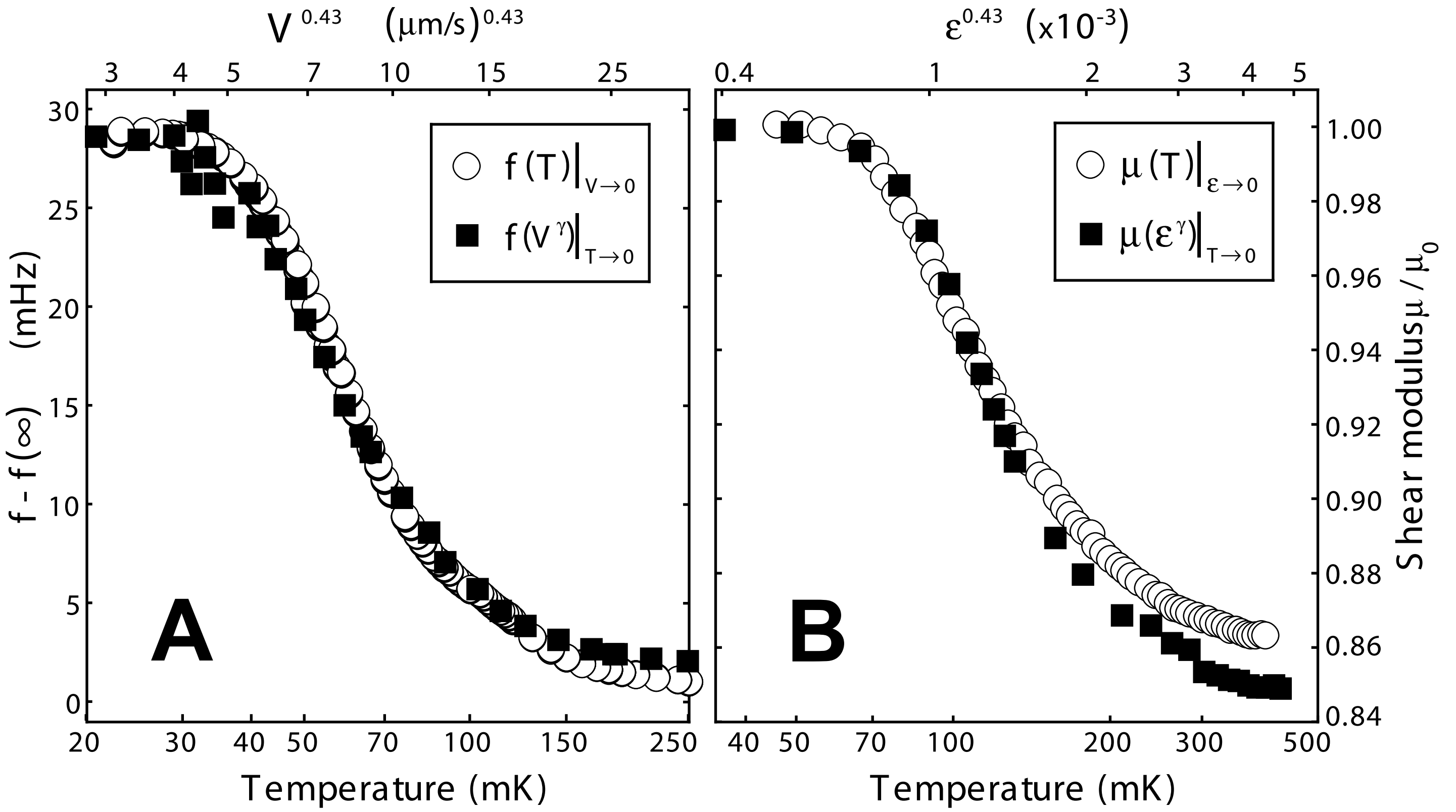

**Supplementary Online Materials:**

# Interplay of relaxational, rotational and shear dynamics in solid $^4$He

E.J. Pratt, B. Hunt, V. Gadagkar, M. Yamashita, M.J. Graf, A.V. Balatsky, and J.C. Davis

## I Evolution of Relaxation Time-Constant at the Superfluid BKT phase transition

The technique for studying relaxation times for microscopic excitations as introduced in Eqn. 3 of the main text, is demonstrated here for TO data at the well-known BKT phase transition from a normal to superfluid film of $^4$He. In this case the average superfluid velocity $\mathbf{u_s}$ can be written in terms of the normal-fluid/substrate velocity $\mathbf{v}_\mathrm{n}$ and the TO frequency $\omega$ as

$$\mathbf{u_s}(\omega) = [1 - \epsilon^{-1}(\omega)]\mathbf{v}_\mathrm{n}(\omega), \tag{S1}$$

where the dynamical function $\epsilon^{-1}(\omega)$ describes the role of vortex motion in screening the local superfluid velocity field from the substrate motion. In TO experiments, the changes to the dissipation and resonant frequency of the TO due to the superfluid vortex motion can be written as follows:

$$\Delta D = \frac{\rho_s^0 A}{M} \Im[-\epsilon^{-1}(\omega)] \tag{S2}$$

and

$$\frac{2\Delta f_0}{f_0} = \frac{\rho_s^0 A}{M} \Re[-\epsilon^{-1}(\omega)], \tag{S3}$$

where $A$ is the area and $M$ the mass of the substrate, and $\rho_s^0$ is the unrenormalized superfluid density. Taking the ratio $\Delta D/(2\Delta f/f_0)$ directly measures $\Im\epsilon^{-1}(\omega)/\Re\epsilon^{-1}(\omega)$ due to the vortex screening. Figure S1 shows the value of $\Delta D/(2\Delta f/f_0)$ as a function of $T$ calculated from TO measurements of the BKT transition. We see directly that when the phase transition to a long-range superfluid state occurs, there is a dramatic change in this function due to diminishing microscopic relaxation times.

In the main text we demonstrate repeatedly that no such signature is observed in $\Delta D * f_0/2(f_0 - f)$ for solid $^4$He in the temperature range which has been ascribed to a phase transition to a supersolid (Fig. 1 → Fig. 3). Moreover we show that no such

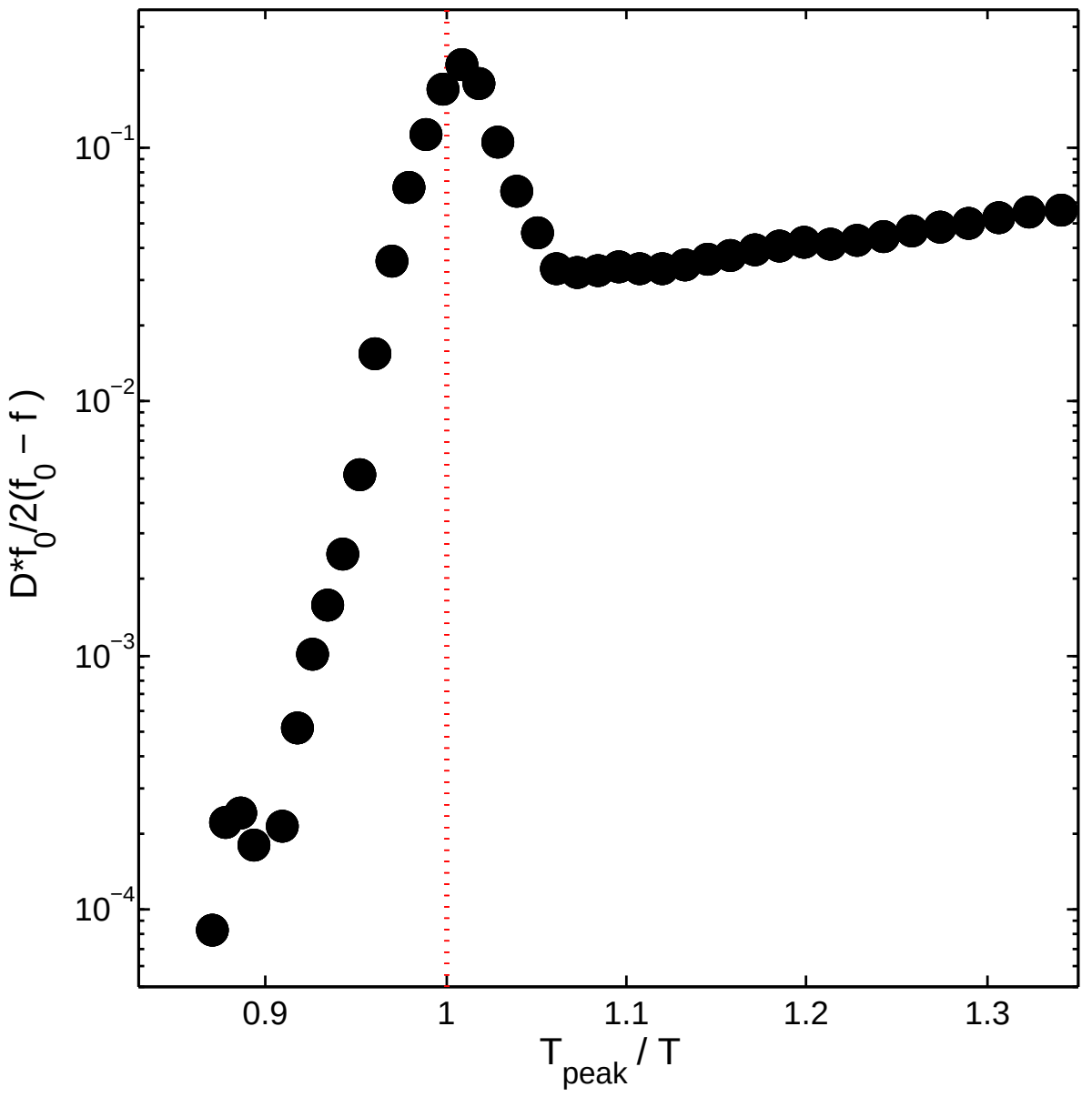

(a) Arrhenius plot using linear $\frac{T_{\mathrm{peak}}}{T}$ axis.

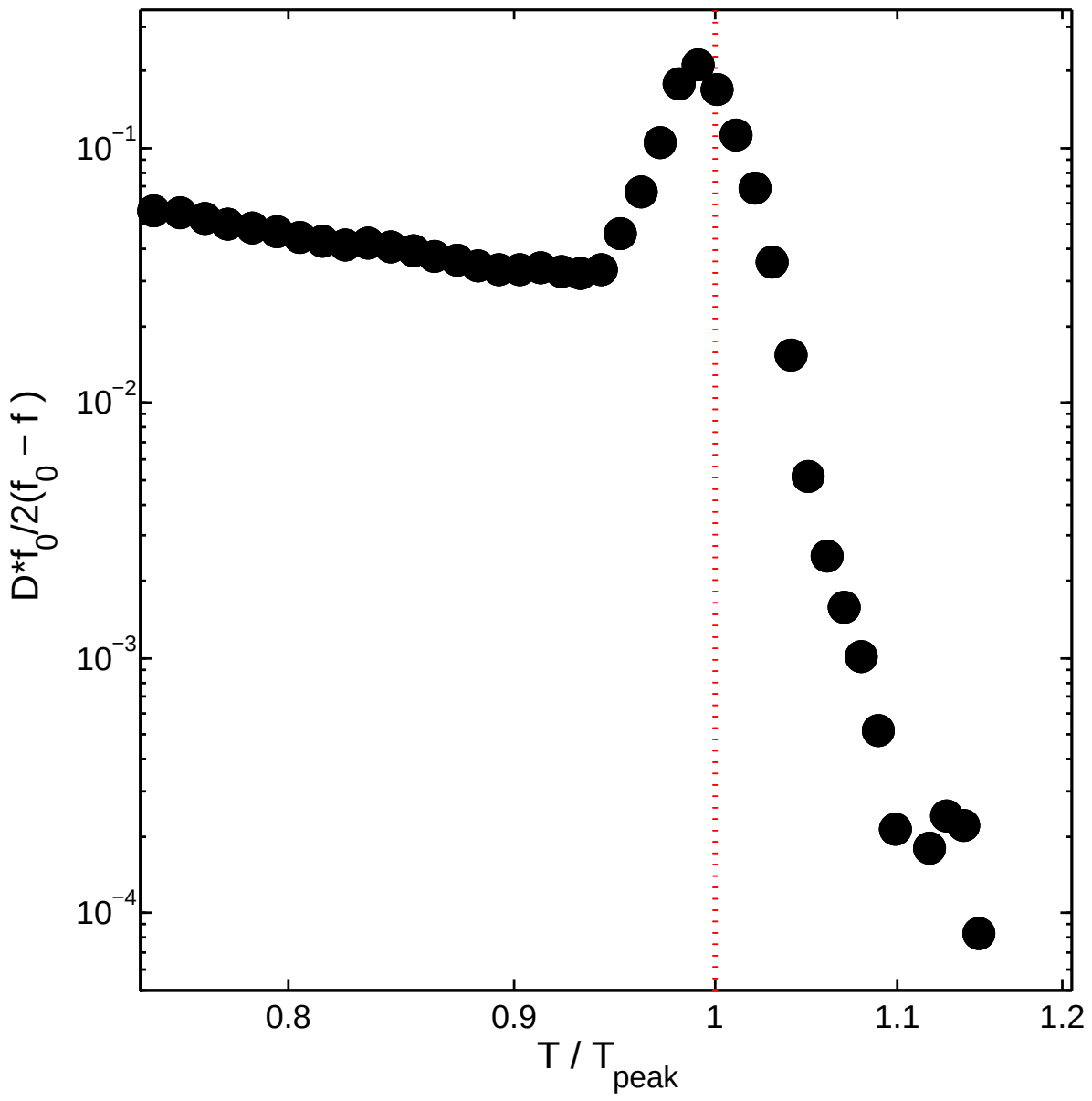

(b) Power-law plot as in the main text; log $\frac{T}{T_{\mathrm{peak}}}$ axis.

Figure S1: Dynamical relaxation time $\frac{Df_0}{2(f_0-f)}$ of thin superfluid film near the BKT transition temperature. $T_{\mathrm{peak}}$ is defined as the temperature at which the dissipation is maximized.

signature in $\Delta D * f_0/2(f_0 - f)$ is observed in the TO rim-velocity range that has been ascribed to the supersolid critical velocity.

# II Materials and Methods

## II.a Sample formation and apparatus

For our studies, solid helium samples are grown from a high-pressure liquid (at 73 bar and 3.3 K) with a nominal $^3$He concentration of 300 ppb by the blocked capillary method, cooling rapidly along the melting curve (approximately 100 minutes from 3.2 K to < 1 K), and they typically reach a low-temperature pressure of 39 bar. The samples are formed inside an annular chamber with a cross-section of 100 microns x 3 mm and radius of 4.5 mm (see Fig. S2A), which corresponds to a surface-to-volume ratio of $200\,cm^{-1}$. The torsion rod is made of annealed beryllium copper and the helium sample chamber is made of Stycast 1266. The resonant frequency at 300 mK is 575.018 Hz for the empty cell and 574.433 Hz for the full cell. The full-cell Q at 300 mK is $4 \times 10^5$.

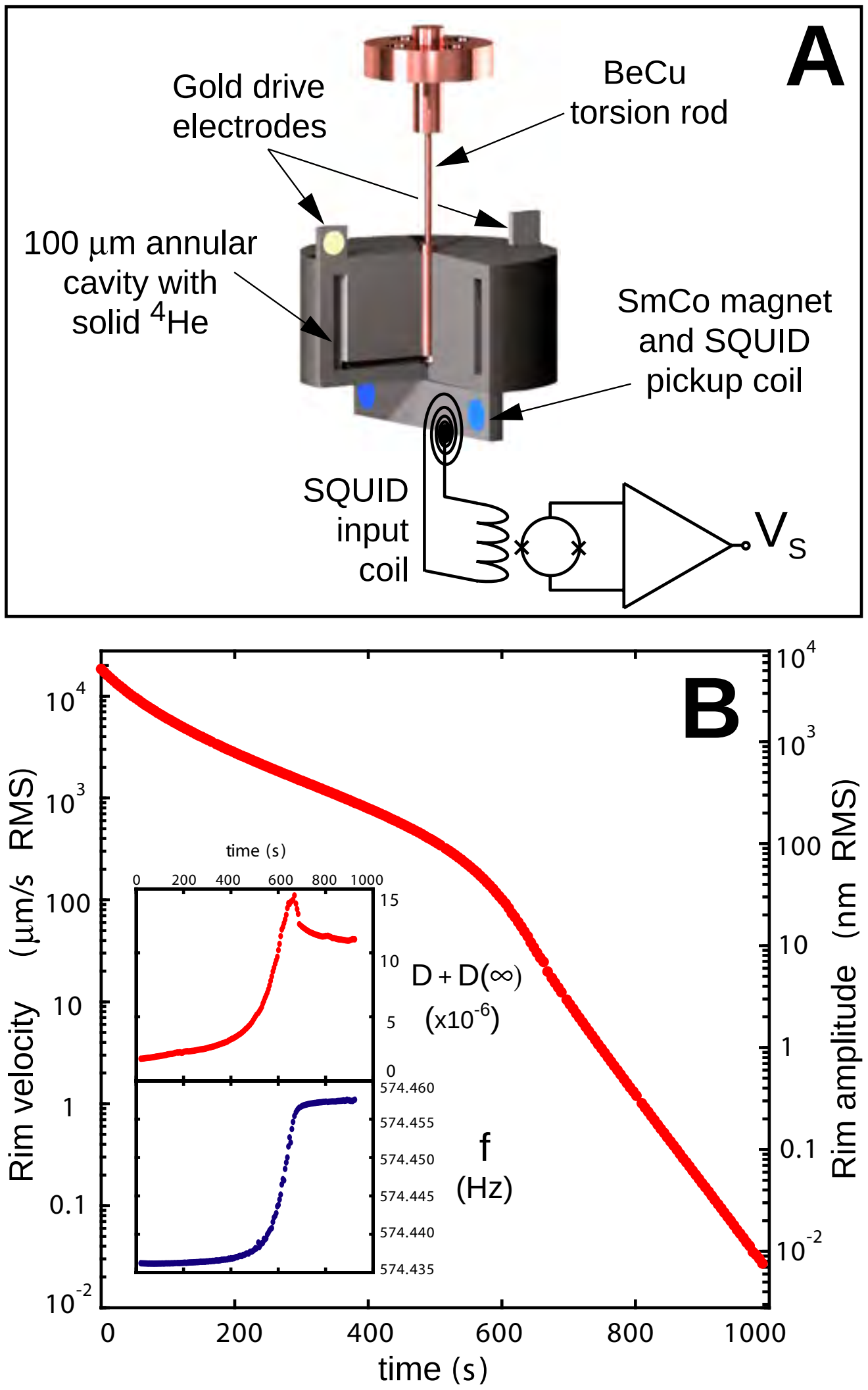

Figure S2: A. Schematic of the SQUID-based torsional oscillator. $^4$He is solidified by blocked-capillary cooling within an annular channel 100 microns wide within a Stycast-1266 cell. The torsional motion can detected either by a capacitor electrode (used at highest velocities) or via the high-sensitivity DC-SQUID detector.
B. A typical FID, taken at 47 mK. This was acquired by stabilizing the temperature, driving the oscillator to high amplitude, and then suddenly turning off the AC drive. The corresponding frequency and dissipation as a function of time during the FID are shown inset.

With sufficiently high signal-to-noise ratio and dynamic range, a TO free inertial decay (FID) in which oscillations of large initial amplitude (or maximum rim-velocity $V$) are allowed to decay freely in time, while both $f(V)$ and the dissipation rate $D(V)$ are measured continuously at fixed temperature, allows complete mapping of the parameter space. Here we introduce such a FID approach to TO studies of solid $^4$He. High-precision sensing of rotational motions is achieved primarily using a SQUID-based detection (Fig. S2A) of the magnetic flux from a permanent magnet attached to the TO body [9]. In Fig. S2B we show a typical FID at a single temperature $T$. From it, the $f(T, V)$ is measured by using a high-precision frequency counter, and $D(T, V)$ is determined by the measured rate of energy loss from the TO. The simultaneously measured $f(T, V)$ and $D(T, V)$ during this FID taken at $T = 47$ mK are shown as insets of Fig. S2B. Our procedure then consists of such measurements over many orders of magnitude in $V$ during each FID, at a closely spaced sequence of different temperatures $T$.

### II.b Rotational susceptibility measurement

Before each FID ringdown, the torsion oscillator is first driven to its initial high-velocity state and thermalized for roughly 330 seconds at the temperature of interest $T_i$. Thermal hysteresis is avoided by these long preparation times at fixed temperature, which are sufficient to relax most of the way towards the asymptotic susceptibility [9]. The ringdown commences by turning off the drive voltage and recording the complex susceptibility as follows: as the oscillator rings down, the resonant frequency is continuously recorded by an Agilent 53131A counter with a 4.0 second integration gate, within a standard phase-locked-loop (PLL) detection circuit.

This recorded resonant frequency $f(T_i, V)$ contains the real part of the solid helium dynamical susceptibility as a function of $V$ (Eqn. 4 of the main text). To extend the dynamic range ever further, at the highest $V$, a calibrated current preamplifier and lockin are used to detect the motion of capacitive electrodes on the T.O. As the velocity decays into the range where the DC-SQUID can lock, the PLL and lockin are switched (within one gate time) to record the motion of a SmCo magnet on the T.O. by a calibrated DC-SQUID detector. Thus, the T.O. velocity is measured directly and continuously recorded during the ringdown.

Later, the quality factor $Q(T_i, V)$ is acquired by curve-fitting the exponential decay constant of the ringdown envelope within a sliding window seven gate-times (28 s) wide. From Eq. 3 of the main text, $D(T_i, V) = 1/Q$ contains the imaginary part of the solid helium rotational susceptibility.

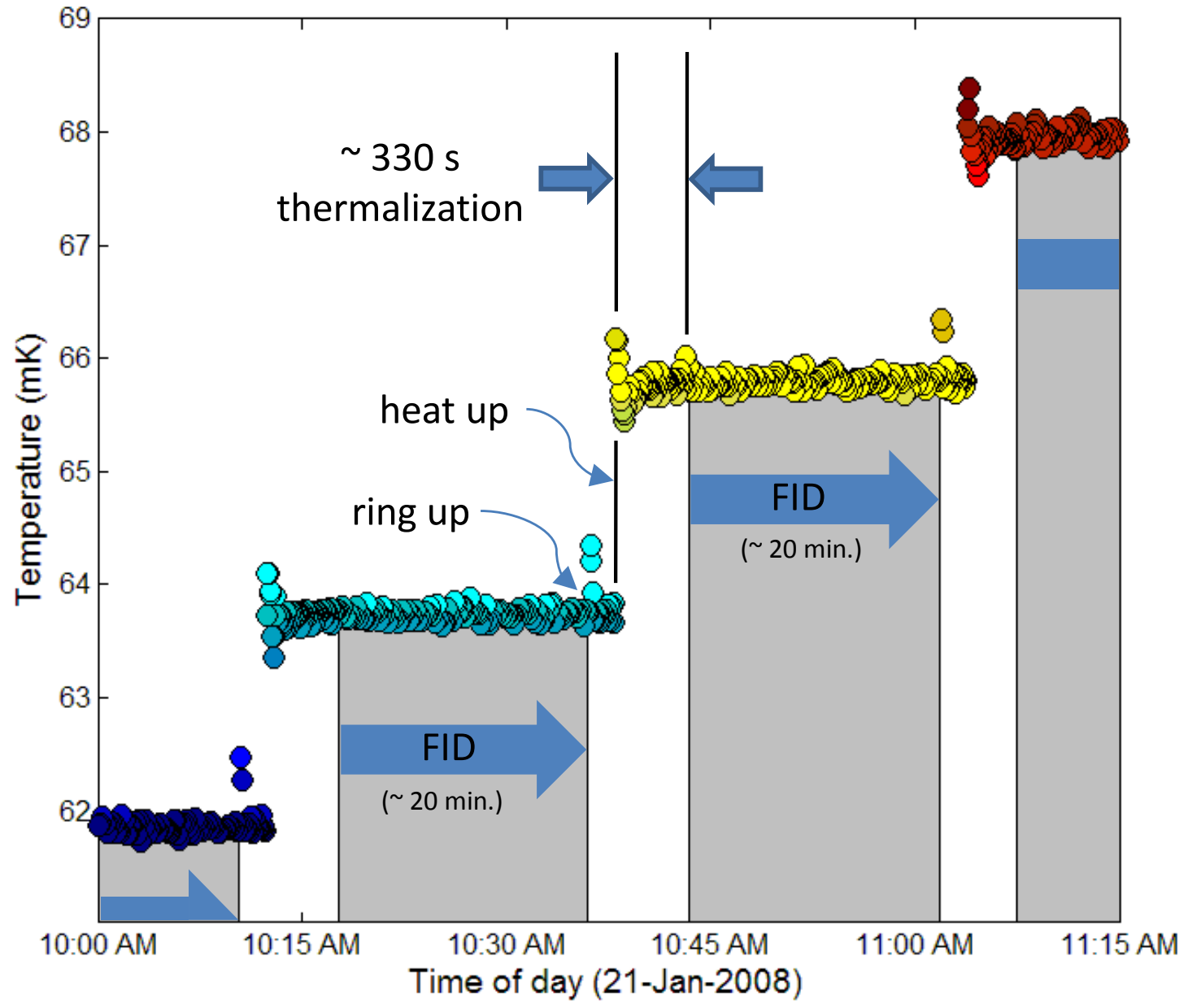

Figure S3: Record of the mixing-chamber temperature during acquisition of the FID susceptibility map.

After the T.O. amplitude decays below the SQUID sensitivity limit (roughly 20 minutes of total FID), the drive voltage is turned back on, the temperature is incremented to $T_{i+1}$, and then the FID procedure is repeated. Each of the two complete sets of 98 FID curves $f(T_i, V)$ and $D(T_i, V)$ with $i = 1 : 98$ are then smoothly interpolated into contoured surfaces as shown in Fig. 1 of the main text. An annotated record of the mixing chamber temperature during the FID-mapping experiments is shown in Fig. S3.

Using this scheme, we can map the susceptibility over more than seven decades of velocity at a data rate of up to two FID per hour, which is fast enough to acquire the full susceptibility in a few days.

### II.c Background subtraction

To extract the dynamical quantities $f(v, T)$ and $D(v, T)$ from FID experiments, the background observations due to non-solid-helium dynamics must be subtracted. To accomplish this, we acquired empty-cell ringdown data which captured the spurious effects of high-amplitude nonlinearity and low-amplitude electronic aliasing in the measured oscillator dissipation and frequency; these data are shown as the black diamonds in Fig. S4. The oscillator dissipation (Fig S4c) is acquired by curve-fitting a sliding exponential window 28 seconds wide across the measured amplitude decay (Fig S4a).

A smooth polynomial curve that captured the shape of the spurious nonlinearities for each dynamical quantity was fitted to these data and offset to correspond at high temperature with the solid helium data; the offsets $f_\infty$ and $D_\infty$ are indicated by the black arrows in Fig. S4b,c. The smoothed curves are indicated by the dashed black lines. After subtraction of these curves, the resultant dynamical quantities $f(v, T)$ and $D(v, T)$ are revealed as reported in the main text. For reference, the 47 mK FID - which was also used for illustration in Fig. S2B - is shown here in Fig. S4 before background subtraction.

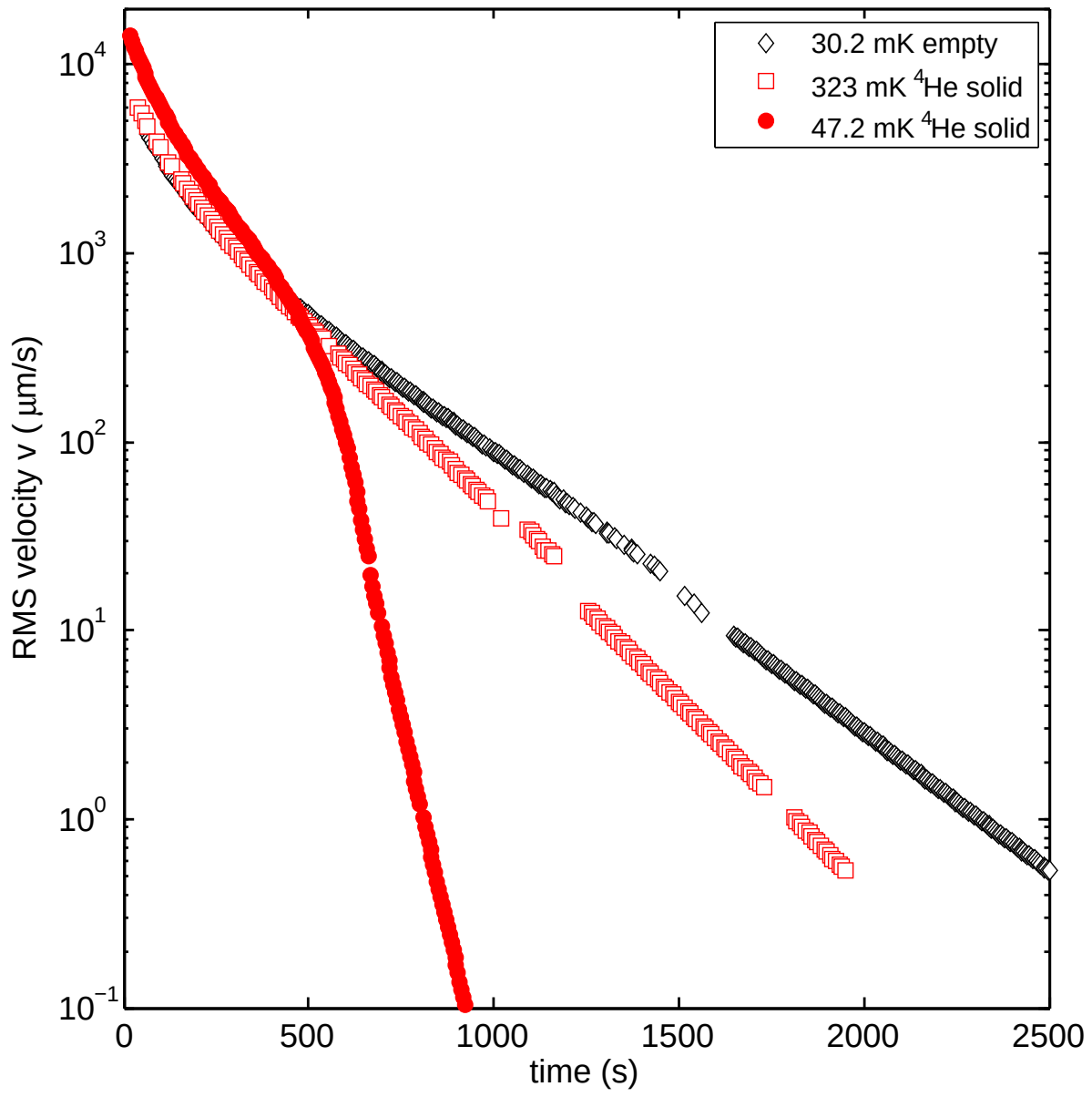

(a) Torsion oscillator amplitude during FID ringdowns.

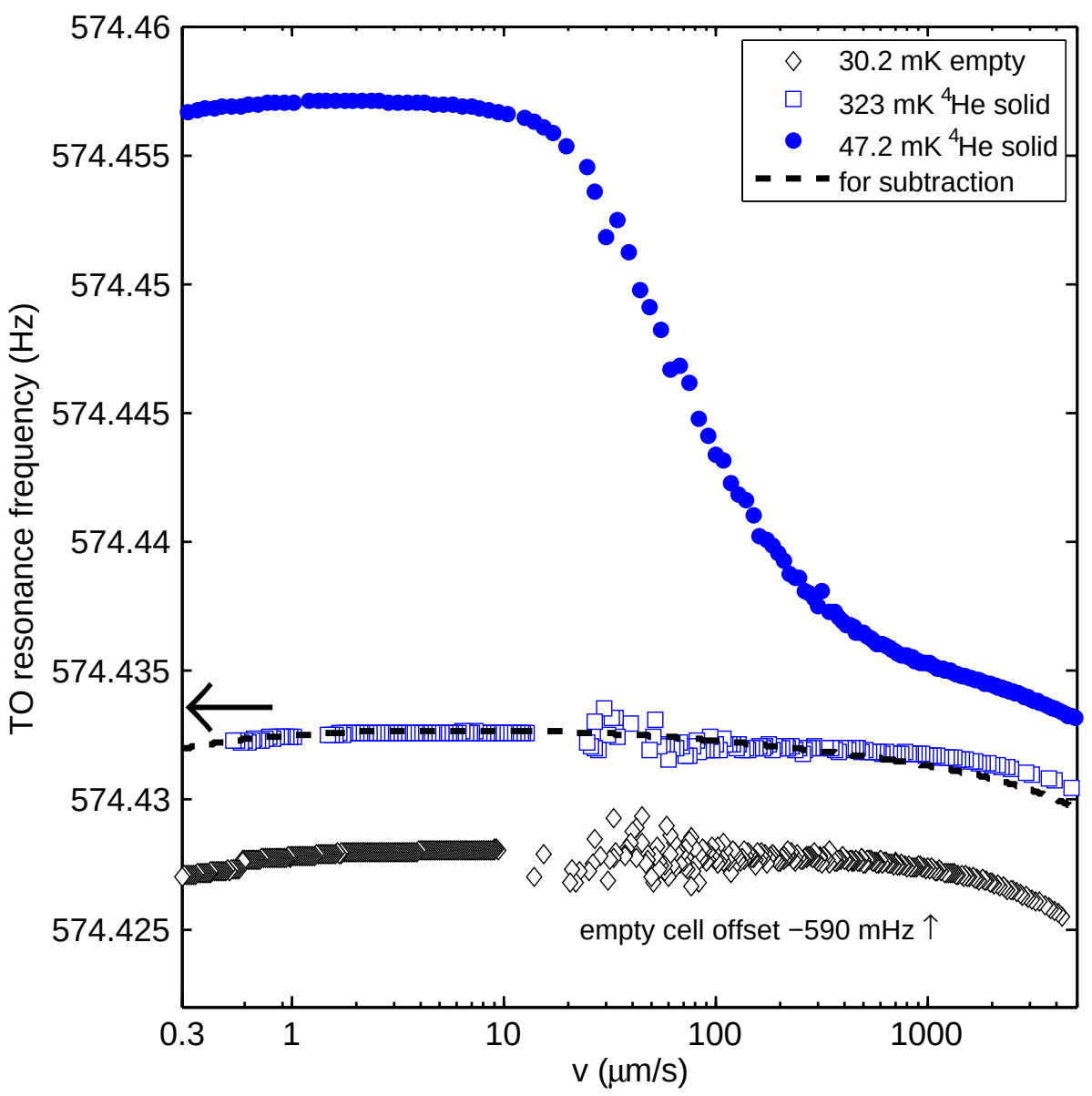

(b) Torsion oscillator resonance frequency. $f_\infty$ is indicated by a black arrow. The empty cell frequency is offset for clarity.

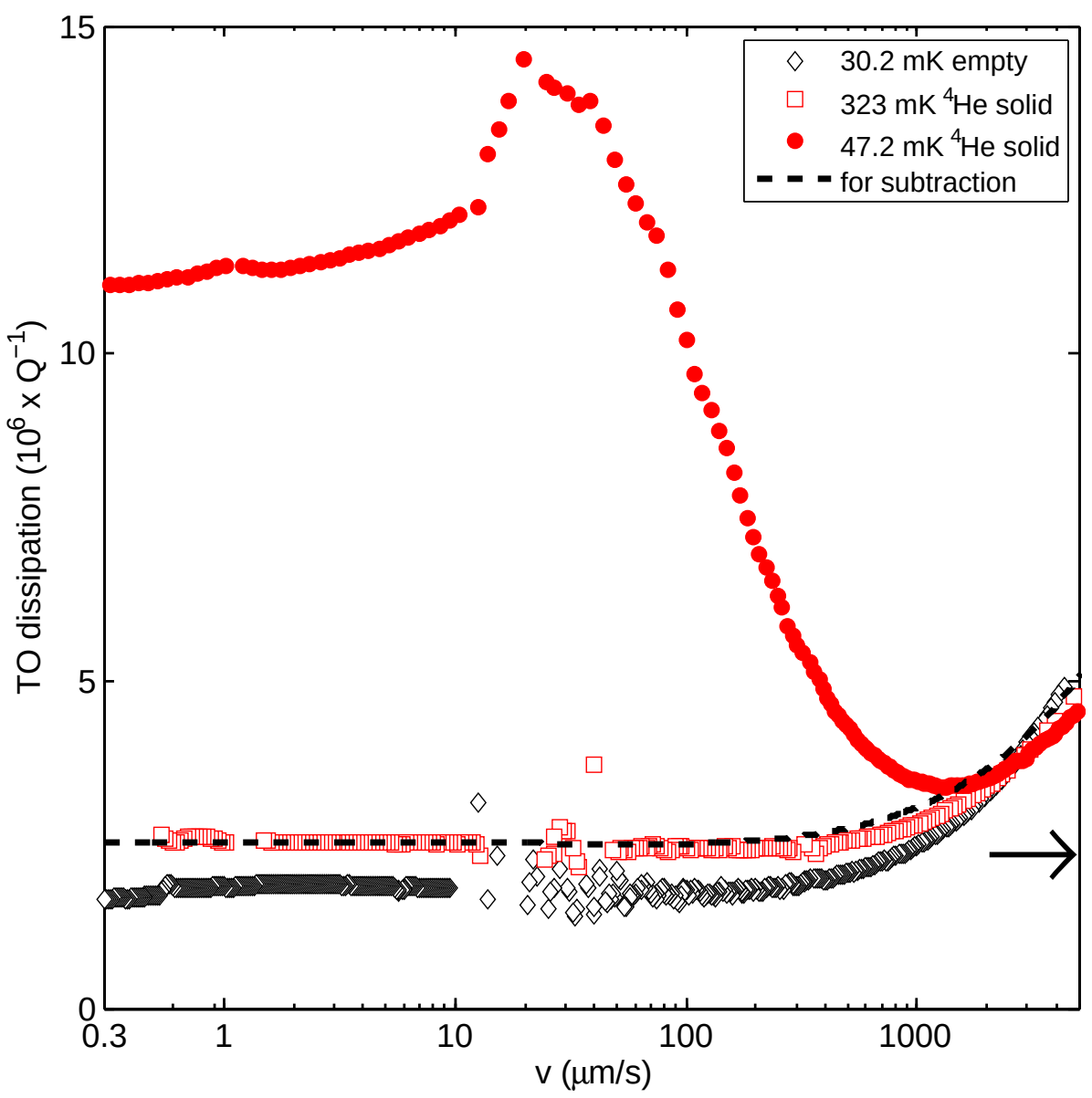

(c) Torsion oscillator dissipation. $D_\infty$ is indicated by a black arrow.

Figure S4: Details of the background surface subtraction from experimental free-inertial-decay data, with empty cell and very high temperature data.

There remains a small, mostly constant (as a function of velocity and temperature) fraction of the total dissipation in the helium - which is visible as the difference between low-velocity empty cell and low-velocity 323 mK solid helium data in Fig. S4 - whose dynamical origin remains unexplained. However, at all temperatures and velocities, the component of the dissipation which we associate with the solid helium dynamics after subtraction of the dashed black line remains by far the dominant observed feature of the oscillator dynamics.

## III Detailed structure of rotational susceptibility data

### III.a Line-cut trajectories on the susceptibility surfaces

The relevant trajectories along which data is extracted from the susceptibility surfaces of Fig. 1A,B of the main text in order to generate Fig. 2A,B,C,D of the main text, are illustrated here in Fig. S5.

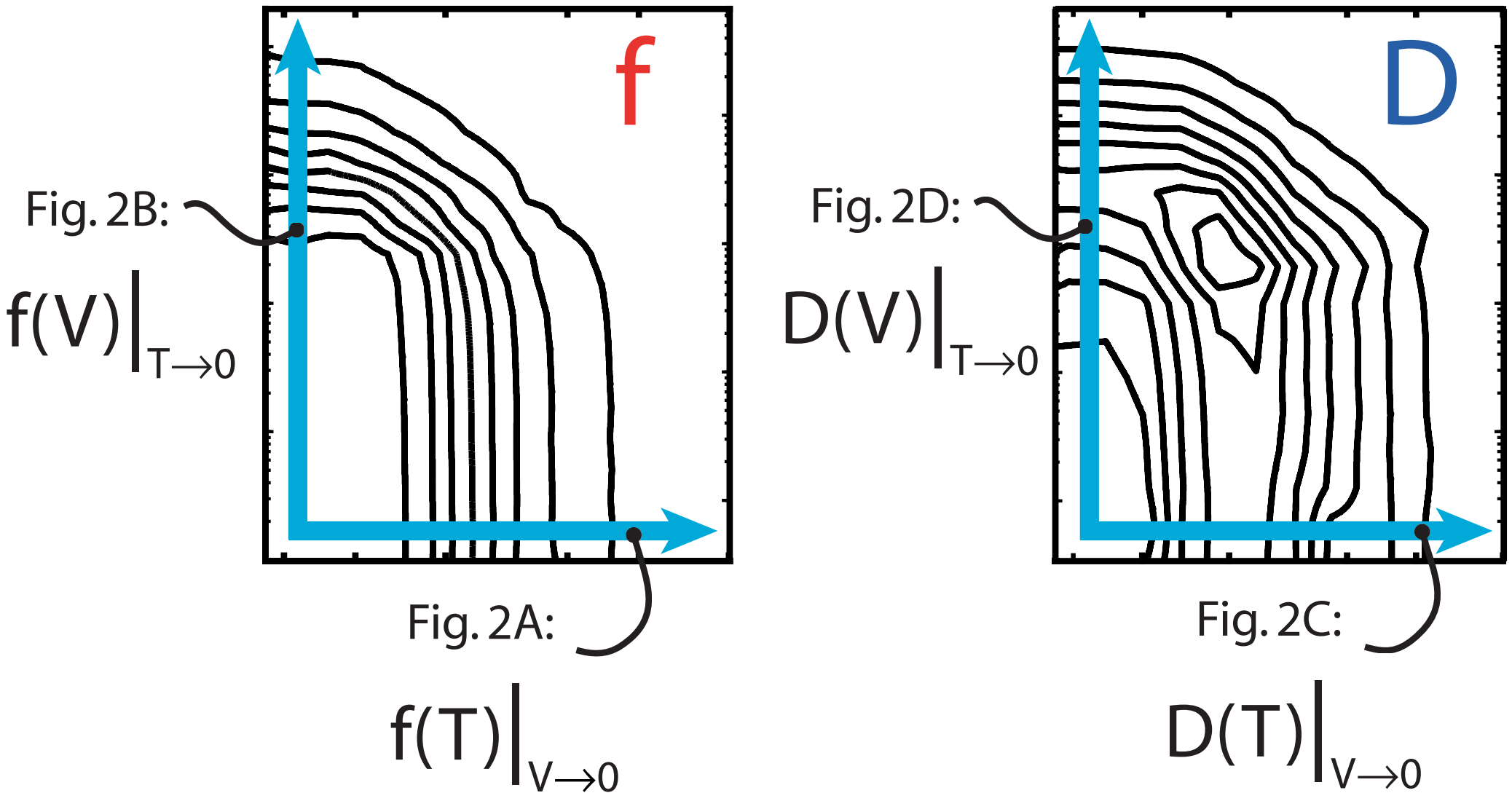

Figure S5: Line cuts across the susceptibility surfaces used for plots in Fig. 2 of the main text. The origin of each panel A-B is indicated by its corresponding trajectory arrow, overplotted on the contoured surfaces representing Fig. 1 of the main text.

### III.b Sharpness of velocity-dependent susceptibility rolloff

The apparent sharpness of the low-velocity rolloff of the frequency shift in Fig. 2B of the main text (near 20 microns/s) might be due to a truly discontinuous slope in the curve $f\left(\sqrt{V}\right)$ and might therefore represent the onset of a true critical velocity at low temperatures. In fact, however, due to switching between different S/N performance regimes with our dual-detector apparatus, at lowest temperatures this shoulder happens to be coincident with a velocity range exhibiting particularly low visibility in the susceptibility.

To see this, consider that a single FID at lowest temperature viewed on a $\sqrt{V}$ axis, as in Fig. 2B of the main text, appears to emphasize the sharpness of the onset of velocity dependence. However, by viewing this data on a log-$V$ axis, we can say that the curves in this temperature range are more consistent with an evolution of smooth-rolloff behavior as exhibited everywhere else (with higher visibility) in the VT plane. This is illustrated in Fig. S6, in which a sequence of velocity-dependent line cuts at various temperatures are displayed.

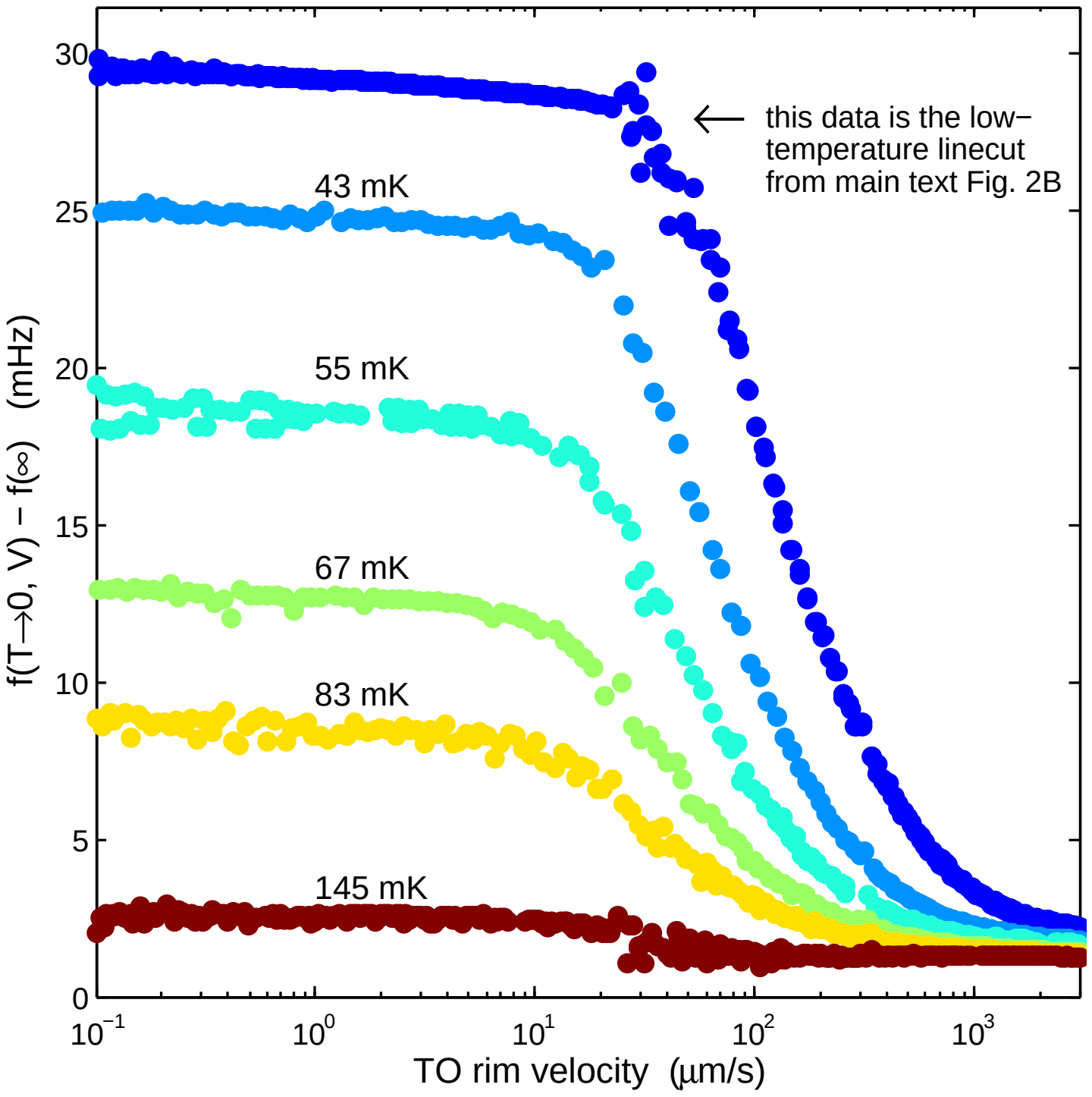

Figure S6: Velocity dependence of the susceptibility at various temperatures. The low-temperature line cut shown here is identical to that in Fig. 2B of the main text, except that this independent axis is logarithmic in $V$ (as opposed to being linear in $\sqrt{V}$.)

Moreover, all of this data is also visible (with no apparent kinks) on the collapsed axis of Fig. 3A of the main text, which is consistent by construction with smooth Debye-like rolloff behavior.

Finally, we note that the lower end of the validity range that we report for power-law scaling behavior illustrated by Fig. 2F of the main text is also nearly coincident with the rolloff feature near 20 microns/s. This is due to the fact that the validity range for the power-law plot is limited by the blow-up of the denominator term $(f_0 - f)$ in the empirical relaxation time $\tau_E$, which becomes extremely small as $f$ approaches $f_0$ at low velocity and renders ambiguous any measurement of $\tau_E$.

Overall, we conclude that we do not observe anything different than consistent power-law scaling outside this velocity range (specifically, 15 to 742 microns/s), but simply that given the available S/N of our apparatus, observations of $\tau_E$ are ambiguous beyond those velocity ranges.

### III.c Absence of sharp features in $\frac{Df_0}{2(f_0-f)}$

To see directly why no sharp feature corresponding to the dissipation peak appears in the relaxation time $\frac{Df_0}{2(f_0-f)}$, it is illustrative to plot both $D$ and $\frac{f_0}{2(f_0-f)}$ on the same axis. This latter quantity is not directly viewed in Fig. 2 of the main text - only the associated quantity $f - f_\infty$ is shown there. Here in Fig. S7 we see the origin of the compensating behavior that eliminates any sharp features after multiplying $D$ by $\frac{f_0}{2(f_0-f)}$.

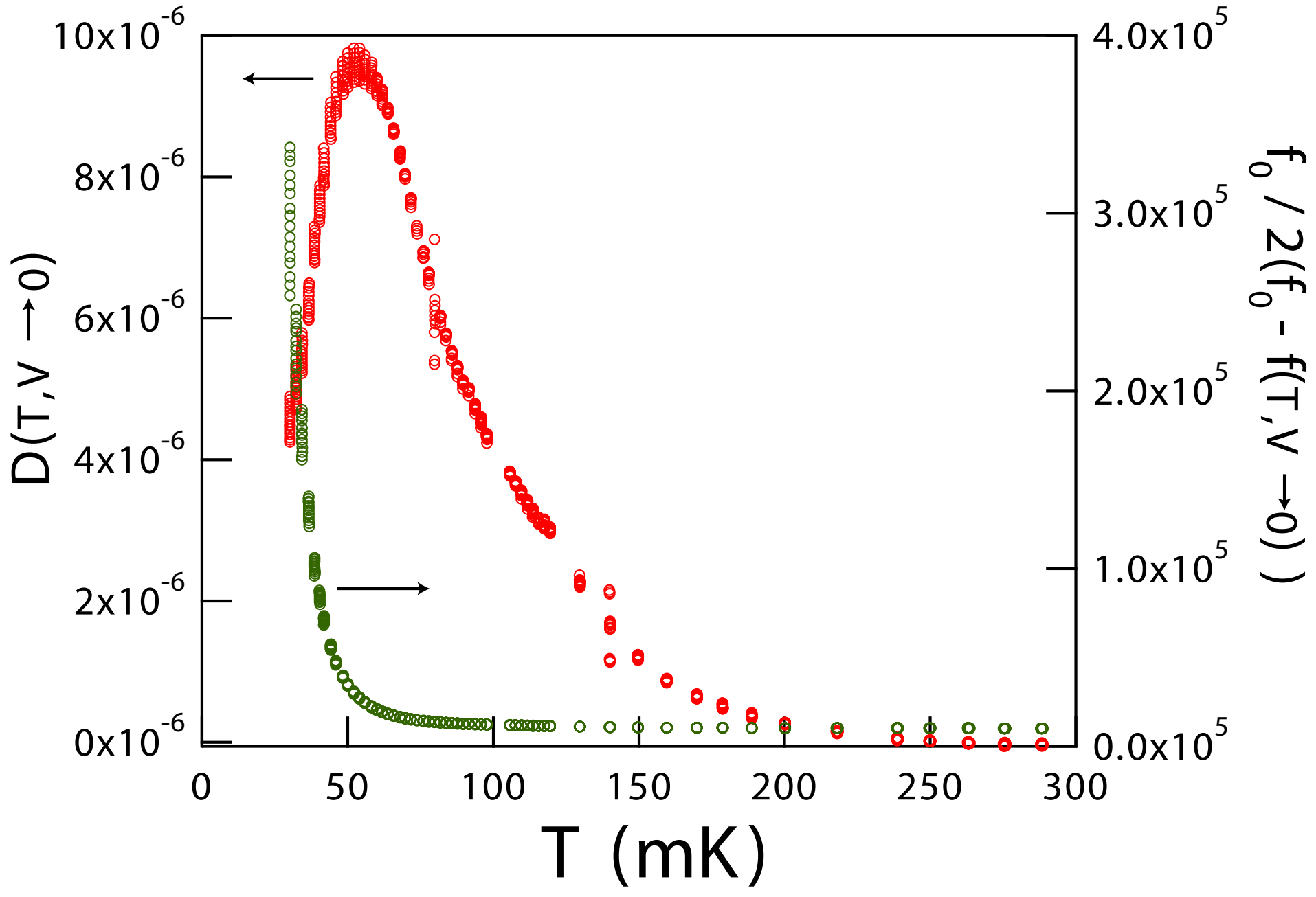

Figure S7: Low-velocity $D$ and $\frac{f_0}{2(f_0-f)}$ measurements plotted independently on the same temperature axis. The product of these two curves has no sharp feature near the temperature of the dissipation peak, and in addition, exhibits a power law behavior.

## IV Unified rotational dynamics

### IV.a Analysis based on the empirical relaxation time $\tau_E(T,V)$

The Debye form $\chi^{-1}_{4\mathrm{He}}(T) = g/(1 - i\omega_0\tau(T))$ discussed in the main text, which captures the phenomenology of a single inertially active mechanical energy barrier, is highly consistent with the near-coincidence of the dissipation peak temperature with the temperature of the frequency shift's maximum slope, as well as the general topology of the curves (Figs 1,2). This choice of back-action, and its natural extension to describe shear agitation, is particularly valuable because it leads to a simple form of relaxation time analysis for experimental data, which is captured by the term $\tau_E(T,V)$ as seen in Eqn. 3 of the main text: $\tau_E(T,V) = \frac{D(T,V)f(0)}{2\omega_0(f(0)-f(T,V))}$.

This empirical relaxation time - though motivated by a particularly simple back-action - is well-defined for any complete experimental set of complex rotational susceptibility observations $f(T,V)$ and $D(T,V)$. However, other more complicated choices for the back-action are also available - which typically seek to describe the inertia of a glassy distribution of mechanical barrier energies - many of which can be optimally fit to the magnitudes of the real and imaginary susceptibilities. One such back-action, which we detail here for illustration and which is also described in [33], reflects the Davidson-Cole distribution function: $\chi^{-1}_{\mathrm{DC}}(T,V) = g/(1 - i\omega_0\tau_{\mathrm{DC}}(T,V))^{\beta}$ with a stretching exponent $0 < \beta \leq 1$ and an associated microscopic time $\tau_{\mathrm{DC}}$. If one used Equation 3 of the main text as an analysis tool for observations of a material which turned out to actually be a Davidson-Cole glass, then one would expect the following relationship between the model parameter $\tau_{\mathrm{DC}}$ (which is not directly observable) and the empirical relaxation time $\tau_E$:

$$\tau_E(T,V) = \frac{D(T,V)f(0)}{2\omega_0(f(0) - f(T,V))} = \omega_0^{-1}\tan\left(\beta\tan^{-1}\left(\omega_0\tau_{\mathrm{DC}}\left(T,V\right)\right)\right) \quad \text{(S4)}$$

In other words, the empirical relaxation time $\tau_E$ exhibited by the experimental data would in fact still be a simple monotonic function of the otherwise inaccessible $\tau_{\mathrm{DC}}$. This illustrates the continued utility of Eqn. 3 and its outcome in Fig. 3 of the main text as analytical tools for a torsion oscillator containing a material of possibly more complex rotational susceptibility, in the sense that a sudden change in relaxation rates at a $T_C$ and/or $V_C$ of a superfluid would still be detectable through measurements of $\tau_E$.

### IV.b Unifying the two contributions to the excitation rate

To develop a combined temperature-shear agitation rate, we start by reprinting Equation 5 of the main text here, but normalized via multiplication by $\omega_0^{-1}$:

$$(\omega_0 \tau_E)^{-1} = \frac{(\Sigma/\omega_0)}{T^\zeta} + \frac{(\Lambda/\omega_0)}{V^\lambda} \tag{S5}$$

and if we assume that the unknown fit parameters $\Sigma$ and $\Lambda$ in fact reveal the experimentally relevant quantities $T^\star$ and $V^\star$ by the following power laws:

$$T^\star = \left(\frac{\Sigma}{\omega_0}\right)^{1/\zeta} \text{ and } V^\star = \left(\frac{\Lambda}{\omega_0}\right)^{1/\lambda} \tag{S6}$$

then inserting Eq. S6 into Eq. S5 gives the following unified relaxation rate:

$$(\omega_0 \tau_E)^{-1} = \left(\frac{T^\star}{T}\right)^\zeta + \left(\frac{V^\star}{V}\right)^\lambda \tag{S7}$$

which is the definition of the independent variable axis used to collapse the FID surfaces into Fig. 3A of the main text.

The characteristic values $T* = 63.0\,\text{mK}$ and $V* = 148\,\mu\text{m/s}$ are defined in the main text to be the values which extinguish 50% of the total frequency shift (at low $V$ and low $T$, respectively) as shown in Fig 2A,B of the main text. The exponents $\zeta = -2.75$ and $\lambda = -1.17$ are extracted from power-law fits to $\omega_0 \tau_E$ (at low $V$ and low $T$, respectively), as shown in Fig 2E,F of the main text.

## V Relationship between Shear and TO Experiments

There are two apparent difficulties with the hypothesis that the TO rim-velocity is not the relevant physical quantity and instead it is the acceleration-induced inertial shear strain $\varepsilon$ which is key. First, the inertial shear required to extinguish the frequency increase $f(V)\,|_{T\to 0}$ is estimated (but not measured) to be two orders of magnitude smaller than the direct shear required to equivalently soften the modulus $\mu(\varepsilon)\,|_{T\to 0}$. But no contradiction may exist here because, for a given inertial shear strain in a TO, all the resultant excitations are confined within the annular sample where they unavoidably affect the inertia, while in the direct shear cell the excitations can propagate away through the open edges of the drive plates. Thus it may not be surprising that much greater direct shear agitation is required to achieve the same effective population density of excitations as in the TO. Second, the $\mu(T)$ measured in the direct shear study does not stiffen sufficiently to account theoretically for the full TO frequency shift of an elastic material. Again this may not be a true contradiction because a shearing experiment does not sense the inertia due to any DC mass flow, while the TO frequency shift should be highly sensitive to it. Regardless, since it has not been possible to carry out simultaneous high-precision TO and shear measurements (with the shear sensor within the TO annulus containing solid $^4$He which is undergoing rotation), it is unresolved if these conceptual concerns have physical validity.